\documentclass[twocolumn]{aastex631}
\shorttitle{Radio and X-ray emission in AGNs}
\shortauthors{Wu et al.}
\graphicspath{{./}}

\usepackage{amsmath}
\usepackage{extarrows}
\usepackage{romannum}
\usepackage{multirow}
\usepackage{xcolor}
\begin{document}
\title{A Shock-based Interpretation of Radio and X-ray Emission in Active Galactic Nuclei}

\author[0000-0002-6292-057X]{Fan Wu}
\affiliation{School of Physics and Astronomy, Yunnan University, Kunming 650091, P. R. China\\}
\affiliation{Key Laboratory of Astroparticle Physics of Yunnan Province, Yunnan University, Kunming 650091, P. R. China\\}

\author[0000-0001-7908-4996]{Benzhong Dai}\email{bzhdai@ynu.edu.cn}
\affiliation{School of Physics and Astronomy, Yunnan University, Kunming 650091, P. R. China\\}
\affiliation{Key Laboratory of Astroparticle Physics of Yunnan Province, Yunnan University, Kunming 650091, P. R. China\\}

\begin{abstract}
We propose a shock-based framework to interpret the radio and X-ray emission in active galactic nuclei (AGNs), whose origin remains an open problem. In this framework, the radio emission is produced by synchrotron radiation in the accretion flow or a jet/weak outflow, while the dominant X-ray component depends on the accretion state, the location of the non-thermal emission region, and the available seed photon field. The model provides a self-consistent interpretation of radio and X-ray emission in typical non-jetted AGNs, including low-luminosity AGNs, Seyfert galaxies, and radio-quiet quasars. In jetted AGNs, our results disfavor scenarios in which the non-thermal electrons responsible for X-ray emission are accelerated in the disk or the corona. We use two widely discussed empirical diagnostics, radio loudness and the fundamental plane (FP) of black hole (BH) activity, to assess the applicability and limitations of the model. It can naturally explain the observed trends that the radio loudness increases with BH mass and decreases with the Eddington ratio. The observed slope of the FP depends on how the key physical quantities scale with the accretion rate. As the accretion rate increases, the advection-dominated accretion flow region contracts while the thin disk region expands, reflecting a transition toward a more radiatively efficient accretion structure. The Eddington ratio therefore influences the accretion structure, and may in turn shape the observed AGN classes.

\end{abstract}

\section{Introduction\label{sec1}}

The non-thermal emission of active galactic nuclei (AGNs) mainly arises from accretion flow or jet/outflow in the vicinity of a black hole (BH). Accretion around the BH can proceed either through a radiatively efficient accretion flow (a thin disk with a corona), or through a radiatively inefficient accretion flow (an advection-dominated accretion flow, ADAF). A hot ($T \simeq 10^9\ \mathrm{K}$), optically thin corona can produce X-ray emission, and may represent a continuation of the inner ADAF \citep{1997ApJ...489..865E,2014ARA&A..52..529Y}. However, the origin and detailed structure of the corona remain uncertain. Although the corona is expected to contain a substantial population of electrons in thermal equilibrium, a non-thermal electron component may also make a non-negligible contribution to the radiation \citep{2014ARA&A..52..529Y,2017MNRAS.467.2566F,2021A&A...649A..87G,2023Sci...381..961Y,2024MNRAS.527.5627L,2025FrASS..1130392L}.

In addition, emission lines near the BH are broadened, and define the broad-line region (BLR). A ring-like dusty torus (DT) is often present on parsec (pc) scales. The BLR is obscured by dusty gas in some cases, such that the lack of detection of the BLR does not necessarily mean that it is physically absent \citep{1985ApJ...297..621A}.

In this work, we divide AGNs into jetted and non-jetted systems. Jetted AGNs produce relativistic, collimated jets, whereas non-jetted AGNs may still show collimated radio outflows that are generally weaker, slower, and less extended than true relativistic jets \citep{1998MNRAS.299..165B,2006A&A...455..161L,2004A&A...417..925M,2017NatAs...1E.194P}.

Radio and X-ray emission are two main probes of non-thermal processes operating in AGNs. Radio emission arises mainly from synchrotron radiation produced by relativistic electrons in the jet, hot flow, and disk wind/outflow. X-ray emission is usually produced by inverse Compton (IC) scattering in the jet, hot flow, and corona \citep{2003MNRAS.345.1057M,2014MNRAS.442..784Z,2016A&A...588A.139M,2019NatAs...3..387P,2020MNRAS.497..482L}.

Specifically, high-resolution radio morphology is the primary diagnostic in the radio band. If the source is resolved on milliarcsecond scales, it is typically identified as a jet from the central engine. If it is resolved at the scale of arcseconds and exhibits a steep spectral slope, it is commonly associated with outflows, photoionization cones, or star-forming regions \citep{1997ARA&A..35..607z,2001ARA&A..39..457K}. In cases where the source remains unresolved even at a milliarcsecond resolution, the radio spectral index $\alpha$ becomes the primary diagnostic. A steep slope ($\alpha> 0.5$) suggests emission from outflows, whereas a flat slope ($\alpha< 0.5$) is indicative of emission originating from the corona or the jet base. 

The origin of X-ray emission is similarly uncertain. X-rays can be produced through IC scattering within the jet in many high-luminosity jetted AGNs. The seed photons may originate from synchrotron radiation within the jet (synchrotron self-Compton, SSC), e.g., \citet{1996ApJ...461..657B, 1985A&A...146..204G}, or from external photon fields associated with the BLR or DT (external Compton, EC), e.g., \citet{2000ApJ...545..107B, 1994ApJ...421..153S, 2009MNRAS.397..985G, 2018MNRAS.477.4749C}. However, \citet{1991ApJ...380L..51H, 1993ApJ...413..507H, 1994ApJ...436..599S, 2014ApJ...784..169J,2019BAAS...51c.126K,2020MNRAS.496..245Z} have suggested that X-ray emission may instead arise from a corona located above the accretion disk. Nonetheless, simulations by \citet{2013ApJ...775..103U} indicate that the corona emission in these simulations does not always provide sufficient power to account for the observed luminosities. From an observational perspective, \citet{2008ApJ...684..811S, 2009A&A...501...89T, 2020MNRAS.497..482L} used systematic analyses of spectral indices, properties of obscuration, and comparisons with more massive AGNs to argue that only part of the X-ray emission can arise from thermal Comptonization in a disk--corona system. One possible reason for this is that these models omit the radiative contribution of non-thermal electrons in the corona. 

Given that even a very small population of high-energy, non-thermal electrons can substantially modify the radiative output \citep{2001MNRAS.325..963W,2011MNRAS.414.3330V}, we focus here on the radiation produced by non-thermal electrons accelerated by shocks. Shocks can arise in disks, coronas, jets, and disk winds/outflows, and have been widely invoked as an important mechanism for particle acceleration in AGNs \citep{2011AA...525A.118I,2014MNRAS.442..784Z,2022Natur.611..677L}.

We first assume that shock acceleration provides the dominant non-thermal electron population in both jetted and non-jetted AGNs, independently of any specific accretion  model. We then examine this framework's applicability and limitations by focusing on the location of the non-thermal emission region and the accretion state. Section \ref{sec:framework} presents the physical framework and its fiducial parameters, while Section \ref{sec3} derives the scalings of the candidate radio and X-ray luminosity. Sections \ref{sec5} and \ref{sec6} then interpret the radio loudness and the fundamental plane of BH activity within this framework. Throughout this paper, physical quantities marked with a prime are measured in the comoving frame of the jet or outflow, while those without a prime are measured in the rest frame of the AGN. Gaussian (cgs) units are used, and we do not consider hadronic models for broadband emission. Unless stated otherwise, we use the term ``emission region'' as a shorthand for the region where shock-accelerated non-thermal electrons produce the radio synchrotron emission and/or X-ray IC emission. Photons from the disk, BLR, and DT are treated as external photon fields.

\section{Physical modeling\label{sec:framework}}

We assume that non-thermal electrons are accelerated by shocks, and follow a power-law distribution with a spectral index that is typical of shock acceleration. In this work, the role of shocks is mainly reflected in the overall modeling framework, the injection scheme, and the particle number density, rather than in a detailed treatment of any specific shock model. This is because such a treatment would introduce many poorly constrained parameters.

\begin{figure*}[t]
\centering
\includegraphics[width=0.95\linewidth]{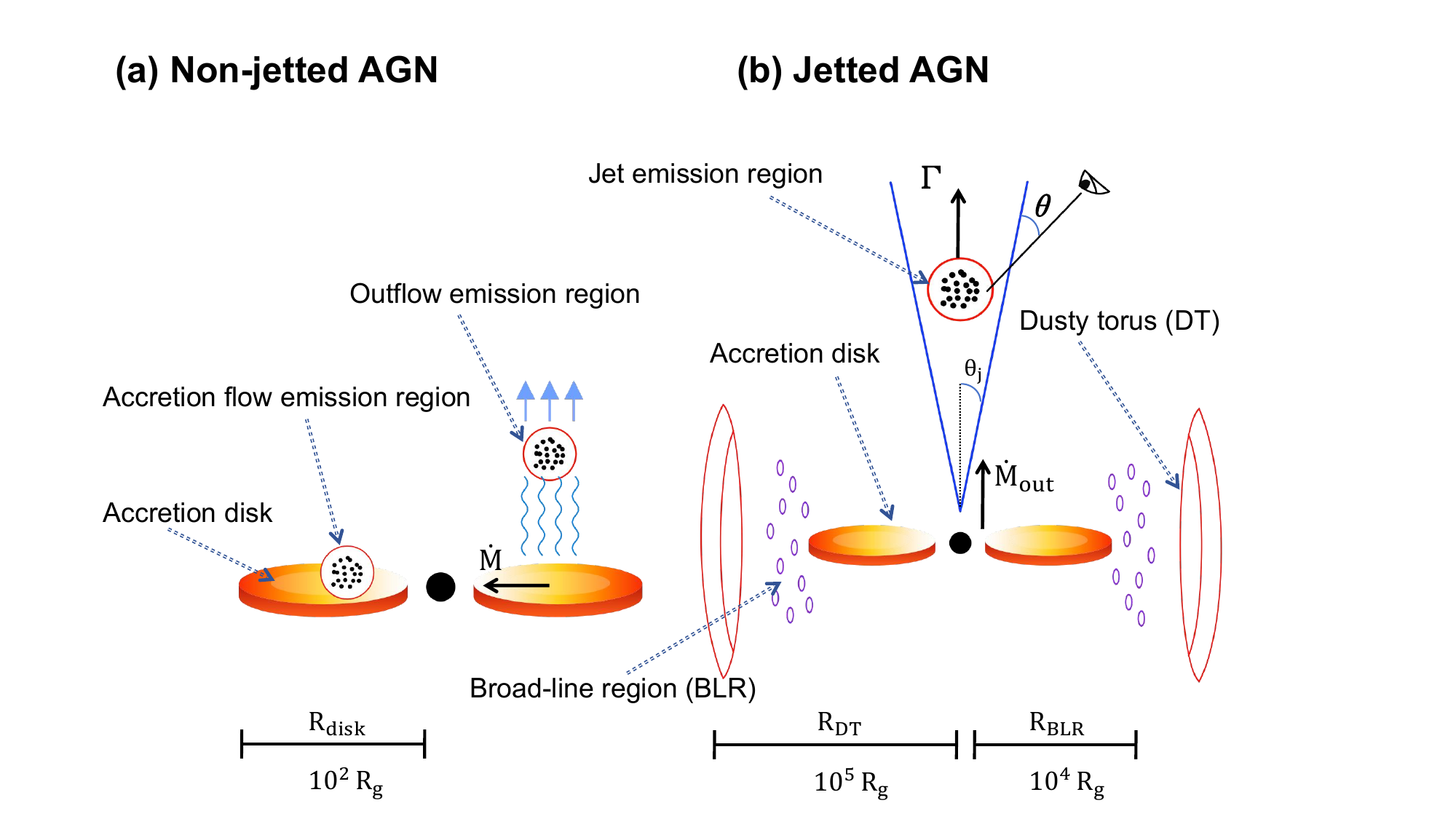}
\caption{Schematic physical setup for the non-jetted and jetted AGN cases. The diagram defines the emission regions, external photon fields, and characteristic radii used in the radio/X-ray luminosity estimates. It is not to scale. \label{fig22}}
\end{figure*}

Figure~\ref{fig22} shows the physical setup considered for non-jetted and jetted AGNs. For non-jetted AGNs, the dominant non-thermal emission is assumed to originate either in the inner accretion flow or in an accretion-driven weak outflow. For jetted AGNs, the emission region is placed in the relativistic jet, where the electrons can scatter synchrotron photons produced in the jet or external photons from the disk and BLR. The characteristic radii adopted for these environments are also indicated, where $R_{\rm g}=2GM_{\rm BH}/c^2$ is the Schwarzschild radius \citep{1989MNRAS.238..897L,1999agnf.book.....K,2002ApJ...570L...9N,2006NewAR..50..728E,2006ApJ...648L.101E,2008MNRAS.387.1669G,2019MNRAS.487.3884C,2020MNRAS.492.5540M}.

Non-thermal electrons accelerated by shocks in or near the emission region are then injected directly into it. For the accretion flow near the BH, a fraction of the accreting material is accelerated and injected into the downstream emission region, in proportion to the inward mass accretion rate $\dot M$ through the inner accretion flow. The particle number density is set by the local residence time. For jets and outflows, we analogously assume that the accelerated particles come from the outflowing material, with their injection rate proportional to the mass outflow rate $\dot M_{\rm out}$. The number density is also controlled by the residence time. Jet and outflow powers are generally coupled to the accretion power \citep{2008MNRAS.385..283C,2014Natur.515..376G,2014MNRAS.445...81S,2020ApJ...900..125R}. The disk winds and wider-angle outflows can also carry away a substantial fraction of the accreting gas and modify the radial accretion rate \citep{2012MNRAS.426..656S,2015Natur.519..436T,2017A&A...601A.143F,2016MNRAS.459..585E,2020MNRAS.494.3616N}. We therefore introduce a ratio of the mass outflow rate to the accretion rate \citep{2008MNRAS.385..283C}, $\mathcal{F}\equiv  \dot M_{\rm out}/\dot M $.

When inferring the geometry of the shock and the emission region, the spectra of AGNs alone do not contain direct geometrical information. Instead, their spectra are the convolution of a source of radiation and the line-of-sight optical depth \citep{2015ARA&A..53..365N}. Thus, we follow the blob model for jetted AGNs: The emission region inside the jet is treated as a spherical zone in the comoving frame, while the region as a whole moves along the jet with a bulk Lorentz factor $\Gamma$, and contains an isotropic magnetic field $B'$. The electron Lorentz factors lie between $\gamma'_{\rm min}$ and $\gamma'_{\rm max}$ in the comoving frame.

We assume an injected electron spectrum produced by shock acceleration, $ Q(\gamma)=Q_0\,\gamma^{-p}$, for $\gamma_{\min}<\gamma<\gamma_{\max}$.
The steady-state equation of continuity in the emission region is
\begin{equation}
\frac{d}{d\gamma}\!\big(\dot\gamma\,N(\gamma)\big)+\frac{N(\gamma)}{t_{\rm esc}} \;=\; Q(\gamma),
\end{equation}
where $\dot\gamma$ is the rate of energy loss owing to synchrotron and IC cooling, and $t_{\rm esc}$ is the escape timescale. For advective (convective) escape, the characteristic timescale satisfies $t_{\rm esc}\ \gtrsim\ \frac{R}{\beta c}$, with $\beta\le 1$ as the bulk speed.

If the band of interest lies in the slow cooling limit, the cooling timescales are $t_{\rm cool}^{-1} \ll t_{\rm esc}^{-1}$, and escape is energy independent. The steady-state electron distribution can then be approximated as
\begin{equation}
N(\gamma)\ \approx\ Q(\gamma)\,t_{\rm esc}
\;=\; (Q_0\,t_{\rm esc})\,\gamma^{-p}
\;\equiv\; N_0\,\gamma^{-p},
\end{equation}
where $N_0$ is the normalization of the number distribution of electrons, and $p$ is the power-law index of their distribution function. 

For the accretion flow case, we assume that a fraction $\eta=5\%$ \citep{1989A&A...224...10C,2011AA...525A.118I} of the accreted electrons participates in the acceleration and radiative processes. We therefore set the normalization to $N_0 =\eta \langle N_e \rangle$, where $\langle N_e \rangle = \dot{M}t_{\rm res}/m_{\rm p}$ is the average number of electrons in the emission region. We take the characteristic residence time to be the local free-fall time, $t_{\rm res} \simeq t_{\rm dyn} =  \sqrt{R^3/GM_{\rm BH}}$. For the jet/outflow case, we assume that the injected material is initially an electron--proton plasma. The number of radiating leptons may be increased by $e^{\pm}$ pairs produced through $\gamma\gamma$ absorption. We parametrize this possible pair contribution by $\mathcal{M}$ and write $N_0 =(1+\mathcal{M})\langle N_e \rangle =  (1+\mathcal{M})\dot{M}_{\rm out}t_{\rm res}'/(\Gamma m_{\rm p}) $, assuming free escape with $t_{\rm res}' \simeq  t_{\rm esc}' =  R_{\rm blob}/c $, where $R_{\rm blob}$ is the size of the emission region. Since we adopt zero pair content, we set $\mathcal{M}=0$. 

Redshift correction and the relativistic Doppler effect must be considered in the observer's frame. Considering the angle $\theta$ between the jet and the line of sight, the Doppler factor is $D =  1/(\Gamma (1 - \beta \cos \theta))$. It approaches $1/\Gamma$ for $\theta\simeq90^\circ$ and $2\Gamma$ for $\theta\simeq0^\circ$ in the relativistic limit $\beta\to1$. The size of the emission region is constrained by variability, as $R_{\rm blob} \lesssim  cDt_{\rm min,var}/(1 + z) $. Here $t_{\rm min,var}$ is the observed minimum variability timescale in the relevant band \citep{2024ApJ...971...74E,2023AA...678L...4P}. For an order-of-magnitude estimate, we adopt a representative few-hour value, $t_{\rm min,var}=10^4\,{\rm s}$.

Assuming a conical jet with a half-opening angle $\theta_j \simeq \frac{1}{\Gamma}$, the distance between the emission region and the central BH is $R  \approx R_{\rm blob} / \theta_j \lesssim  2c \Gamma^2 t_{\rm min,var}/(1+z) $. This sets a lower limit on the bulk Lorentz factor: $\Gamma_{\rm min} \simeq \sqrt{ R(1+z)/(2c  t_{\rm min,var})}$. The minimum bulk Lorentz factor $\Gamma_{\rm min}$ should be
\begin{equation}
\begin{aligned}
 2 \left(\frac{R}{10^2 \  R_{\rm g} }\right)^{\frac{1}{2}} \left(\frac{M_{\rm BH}}{10^8 M_\odot  }\right)^{\frac{1}{2}}\left(\frac{t_{\rm min,var}}{10^4 \ \mathrm{s}}\right)^{-\frac{1}{2}}\left(1+z\right)^{\frac{1}{2}},
\end{aligned}
\end{equation}
for non-jetted AGNs, and
\begin{equation}
\begin{aligned}
 22 \left(\frac{R}{10^4  \  R_{\rm g}}\right)^{\frac{1}{2}} \left(\frac{M_{\rm BH}}{10^8 M_\odot  }\right)^{\frac{1}{2}}\left(\frac{t_{\rm min,var}}{10^4 \ \mathrm{s}}\right)^{-\frac{1}{2}}\left(1+z\right)^{\frac{1}{2}},
\end{aligned}
\end{equation}
for jetted AGNs.

In the following discussion, we adopt representative values of $\Gamma\sim 1$ for non-jetted AGNs and $\Gamma\sim 20$ for jetted AGNs. Finally, for the subsequent calculations, we assume that the size of the emission region is comparable to its distance from the BH, i.e., $R_{\rm blob} \simeq R$. This is reasonable, given the modest bulk Lorentz factors and the fact that regions located closer to the BH tend to be more compact.

\begin{deluxetable*}{lllcl}[!ht]
\tablewidth{0pt}
\movetableright=0cm
\tablecaption{Accretion-rate ranges inferred from the observed radio luminosities of different AGN classes while assuming a typical BH mass of $M_{\rm BH}=10^8\,M_\odot$. Column 1 lists the AGN class, Column 2 gives the corresponding range of radio luminosity, Column 3 gives the assumed radio origin, Column 4 gives the range of $\dot M$ for non-jetted AGNs and $\dot M D^{7/2}$ for jetted AGNs, and Column 5 provides a brief description of the associated physical implications. \label{tab:radio_mdot}}
\tablehead{
\colhead{Source class} &
\colhead{\begin{tabular}{@{}c@{}}Radio luminosity\\[-0.4ex]
\scriptsize $({\rm erg\,s^{-1}})$\end{tabular}} &
\colhead{\begin{tabular}{@{}c@{}}Assumed radio\\[-0.4ex]
origin\end{tabular}} &
\colhead{\begin{tabular}{@{}c@{}}Inferred accretion rate\\[-0.4ex]
\scriptsize $(M_\odot\,{\rm yr^{-1}})$\end{tabular}} &
\colhead{Physical implication}
}
\startdata
\parbox[t]{0.16\textwidth}{Non-jetted AGNs} &
\parbox[t]{0.18\textwidth}{$10^{35}$--$10^{41}$ \citep{2001ApJS..133...77H,2019MNRAS.482.5513L}} &
accretion flow &
$10^{-6}$--$1$ &
\parbox[t]{0.22\textwidth}{The inferred $\dot M$ range includes typical values.} \\
 &  &
weak outflow &
$ 10^{-2}$--$10^{4}$ &
 \\
\hline
\parbox[t]{0.16\textwidth}{Jetted AGNs} &
\parbox[t]{0.18\textwidth}{$10^{39}$--$10^{45}$ \citep{2012MNRAS.421.1569B,2024MNRAS.529.3699Z}} &
jet &
$10^{-4}$--$10^{2}$ &
\parbox[t]{0.22\textwidth}{$D$ cannot be disentangled from $\dot M$.} \\
\enddata
\end{deluxetable*}

\section{Radio and X-ray emission\label{sec3}}

We consider possible sites for the origin of radio and X-ray emission. Radio emission is assigned to synchrotron radiation from either the accretion flow, the jet ($\mathcal{F} \simeq 10^{-1}$), or the weak outflow ($\mathcal{F} \simeq 10^{-4}$). We consider three seed-photon fields for the X-ray emission: synchrotron photons produced locally in the emission region (SSC/ADAF case), thermal photons from the accretion disk (disk/corona case), and reprocessed photons from the BLR (EC-BLR case). We present the luminosity scalings here. The detailed derivations are given in Appendices \ref{app:1}--\ref{app:2}.

\subsection{Radio synchrotron luminosity \label{sec3.1}}

Under the assumptions of our model, radio luminosity in the observer's frame can be written separately for synchrotron emission produced by shocks in the accretion flow and for that produced by shocks in the jet/outflow as follows:

\begin{widetext}
\begin{equation}
\begin{aligned}
L_{\rm R,flow,obs}
&\simeq
2.77 \times 10^{41}
\frac{D^{\frac{7}{2}}}{(1+z)^{\frac{3}{2}}} 
\left( \frac{B}{0.1\ {\rm G}} \right)^{\frac{3}{2}}
\left( \frac{\eta}{0.05} \right)
\left( \frac{R}{10^{2} \ R_{\rm g}} \right)^{\frac{3}{2}} 
\left( \frac{\dot{M}}{M_\odot\,{\rm yr}^{-1}} \right)
\left( \frac{M_{\rm BH}}{10^{8} M_\odot } \right)
\left( \frac{\nu_{\rm c,syn,obs}}{10^{14}\  \mathrm{Hz}} \right)^{\frac{1}{2}}
  {\rm erg\ s^{-1}},
\label{eq:1}
\end{aligned}
\end{equation}

\begin{equation}
\begin{aligned}
L_{\rm R,jet,obs}
&\simeq
1.96 \times 10^{43}
\frac{D^{\frac{7}{2}}}{(1+z)^{\frac{3}{2}}}\left(
\frac{\Gamma}{20}
\right)^{-1} 
\left( \frac{B'}{0.1\ \mathrm{G}} \right)^{\frac{3}{2}}
\left(
\frac{\mathcal{F}}{0.1}
\right)
\left( \frac{R'}{10^{4}\,R_{\rm g}} \right)  
\left( \frac{\dot{M}}{M_\odot\ {\rm yr}^{-1} } \right)
\left( \frac{M_{\rm BH}}{10^{8} M_\odot } \right)
\left( \frac{\nu_{\rm c,syn,obs}}{10^{18}\ \mathrm{Hz}} \right)^{\frac{1}{2}}
  {\rm erg\ s^{-1}},
\end{aligned}
\end{equation}

\begin{equation}
\begin{aligned}
L_{\rm R,outflow,obs}
&\simeq
3.92 \times 10^{37}
\frac{D^{\frac{7}{2}}}{(1+z)^{\frac{3}{2}}}\left(
\frac{\Gamma}{1}
\right)^{-1} 
\left( \frac{B'}{0.1\ \mathrm{G}} \right)^{\frac{3}{2}}
\left(
\frac{\mathcal{F}}{10^{-4}}
\right)
\left( \frac{R'}{10^{2}\,R_{\rm g}} \right) 
\left( \frac{\dot{M}}{M_\odot\ {\rm yr}^{-1} } \right)
\left( \frac{M_{\rm BH}}{10^{8} M_\odot } \right)
\left( \frac{\nu_{\rm c,syn,obs}}{10^{14}\ \mathrm{Hz}} \right)^{\frac{1}{2}}
  {\rm erg\ s^{-1}}.
\label{eq:2}
\end{aligned}
\end{equation}
\end{widetext}

The radio luminosity in the observer's frame follows $L_{\rm R,obs} \propto D^{\frac{7}{2}}$. Doppler beaming can therefore be used to boost the observed luminosity by at most a factor of $\simeq 4\times10^{5}$. This suggests that the observed radio power may be geometric rather than intrinsic. If beaming is neglected and only the intrinsic emission is considered, we obtain $L_{\rm R} \propto \dot M\, M_{\rm BH}$ for both jetted and non-jetted AGNs, which is consistent with the statistical result reported by \citet{2001ApJ...551L..17L}. They found $L_{\rm R} \propto M_{\rm BH}^{2}(L_{\rm disk}/L_{\rm Edd})$, which reduces to $L_{\rm R} \propto \dot M\,M_{\rm BH}$ because $L_{\rm disk} \propto \dot M$ and $L_{\rm Edd} \propto M_{\rm BH}$.

The observed radio luminosities and the corresponding inferred accretion rates are summarized in Table~\ref{tab:radio_mdot}. For non-jetted AGNs, higher inferred $\dot{M}$ values may reflect the contamination of the observed luminosity by non-nuclear emission. For jetted AGNs, very high or very low inferred values of $\dot M$ are more likely to reflect Doppler boosting or de-boosting, rather than genuinely extreme intrinsic accretion rates.

\subsection{X-ray IC luminosity\label{sec4}}

Depending on the origin of the seed photons, IC X-ray emission can be divided into SSC and EC cases. For the SSC case in the observer's frame, we obtain
\begin{widetext}
\begin{equation}
\small
\begin{aligned}
L_{\rm X,ADAF,obs}
&= 5.89 \times 10^{43} \left(\frac{D^3}{1+z}\right) \left(\frac{B}{0.1 \ \mathrm{G}}\right) \left(\frac{\eta}{0.05}\right)^2 \left(\frac{R}{10^2 \  R_{\rm g}}\right)   \left(\frac{\dot{M}}{M_\odot \rm yr^{-1} }\right)^2
\left(\frac{\nu_{\rm c,syn,obs}}{10^{14} \ \mathrm{Hz}}\right)  \ \rm erg \ s^{-1},
\label{eq:3}
\end{aligned}
\end{equation}

\begin{equation}
\small
\begin{aligned}
L_{\rm X,SSC,jet,obs}
&= 2.94 \times 10^{43} \left(\frac{D^3}{1+z}\right) \left(
\frac{\Gamma}{20}
\right)^{-2} \left(\frac{B'}{0.1 \ \mathrm{G}}\right) \left(
\frac{\mathcal{F}}{0.1}
\right)^2     \left(\frac{\dot{M}}{M_\odot \rm yr^{-1} }\right)^2
\left(\frac{\nu_{\rm c,syn,obs}}{10^{18} \ \mathrm{Hz}}\right)   \rm erg \ s^{-1}.
\end{aligned}
\end{equation}
\end{widetext}

Because the external photon field is weak in ADAFs, we consider only the SSC case.

The accretion disk luminosity can be written as
$L_{\rm disk}\simeq \zeta \dot M c^2$, where
$\zeta\equiv L_{\rm disk}/(\dot M c^2)$ is the radiative efficiency. Physically, $\zeta$ measures the fraction of the rest-mass accretion power that is converted into disk radiation. We adopt the fiducial value $\zeta=0.1$, appropriate for a radiatively efficient thin disk \citep{1973A&A....24..337S,2004MNRAS.351..169M,2009ApJ...690...20S,2011ApJ...728...98D,2014ARA&A..52..529Y}. Because the BLR reprocesses a fraction $\xi=0.1$ of the radiation of the disk, its luminosity is $L_{\rm BLR}\simeq \xi \,L_{\rm disk}\simeq \xi \,\zeta \,\dot{M}c^{2}$ \citep{2003ApJ...593..667M}. Here, the corona case refers to the IC scattering of thermal disk photons by non-thermal electrons in the corona, rather than to the conventional thermal Comptonization scenario. In practice, our modeling cannot distinguish between non-thermal electrons originating in the disk and the corona.

In the Thomson regime, the corresponding EC luminosities in the observer's frame are
{\setlength{\abovedisplayskip}{7pt}
\begin{widetext}
\begin{equation}
\small
\begin{aligned}
L_{\rm X,disk,obs}
&= 1.21 \times 10^{48} \frac{D^{\frac{7}{2}}}{(1+z)^{\frac{3}{2}}}  \left(\frac{B}{0.1 \ \mathrm{G}}\right)^{-\frac{1}{2}} \left( \frac{\zeta}{0.1} \right)  \left(\frac{\eta}{0.05}\right)  \left(\frac{R}{10^2 \  R_{\rm g}}\right)^{-\frac{1}{2}}    \left(\frac{\dot{M}}{M_\odot \rm yr^{-1} }\right)^2 \left(\frac{M_{\rm BH}}{10^8 M_{\odot} \ }\right)^{-1}  \left(\frac{\nu_{\rm c,syn,obs}}{10^{14}\ \mathrm{Hz}}\right)^{\frac{1}{2}}  \rm erg \ s^{-1},
\label{eqx}
\end{aligned}
\end{equation}

\begin{equation}
\footnotesize
\begin{aligned}
L_{\rm X,BLR,obs}
= 8.55 \times 10^{44}  \frac{D^{\frac{7}{2}}}{(1+z)^{\frac{3}{2}}} \left(
\frac{\Gamma}{20}
\right)^{-1}  \left(\frac{B'}{0.1 \ \mathrm{G}}\right)^{-\frac{1}{2}} \left( \frac{\xi}{0.1} \right) \left( \frac{\zeta}{0.1} \right)  \left(
\frac{\mathcal{F}}{0.1}
\right) & \left(\frac{R'}{10^4 \  R_{\rm g}}\right)^{-1} \\& \times \left(\frac{\dot{M}}{M_\odot \rm yr^{-1} }\right)^2 \left(\frac{M_{\rm BH}}{10^8 M_{\odot} \ }\right)^{-1}  \left(\frac{\nu_{\rm c,syn,obs}}{10^{18}\ \mathrm{Hz}}\right)^{\frac{1}{2}}   \rm erg \ s^{-1},
\end{aligned}
\end{equation}

\begin{equation}
\small
\begin{aligned}
L_{\rm X,corona,obs}
&=3.82 \times 10^{50}  \frac{D^{\frac{7}{2}}}{(1+z)^{\frac{3}{2}}}  \left(\frac{B}{0.1 \ \mathrm{G}}\right)^{-\frac{1}{2}} \left( \frac{\zeta}{0.1} \right)  \left(\frac{\eta}{0.05}\right) \left(\frac{R}{10 \  R_{\rm g}}\right)^{-\frac{1}{2}}    \left(\frac{\dot{M}}{M_\odot \rm yr^{-1} }\right)^2 \left(\frac{M_{\rm BH}}{10^8 M_{\odot} \ }\right)^{-1}  \left(\frac{\nu_{\rm c,syn,obs}}{10^{18}\ \mathrm{Hz}}\right)^{\frac{1}{2}} \rm erg \ s^{-1}.
\label{eq:4}
\end{aligned}
\end{equation}
\end{widetext}}

Compared with the SSC case, the EC case depends on the BH mass because the seed-photon energy density scales differently in the two cases. When the seed photons are synchrotron photons, their energy density scales as $\propto \dot{M} M_{\rm BH}^{-1}$, whereas it scales as $\propto \dot{M} M_{\rm BH}^{-2}$ for disk photons. As a result, the final luminosity gains an additional factor of $M_{\rm BH}^{-1}$.

\begin{deluxetable*}{lllcl}[!ht]
\tablewidth{0pt}
\movetableright=0cm
\tablecaption{Observed X-ray luminosities and the corresponding inferred constraints on the accretion rate. All accretion rates are estimated by assuming a typical BH mass of $M_{\rm BH}=10^8 \,M_\odot$. Column 1 lists the AGN class, Column 2 gives the corresponding range of X-ray luminosity, Column 3 gives the assumed X-ray origin in our model, Column 4 gives the range of $\dot M$ for non-jetted AGNs, $\dot M D^{7/4}$ for jetted disk/corona and EC-BLR cases, and $\dot M D^{3/2}$ for jetted SSC cases. Column 5 provides a brief description of the associated physical implications. \label{tab:x_mdot}}
\tablehead{
\colhead{Source class} &
\colhead{\begin{tabular}{@{}c@{}}X-ray luminosity\\[-0.4ex]
\scriptsize $({\rm erg\,s^{-1}})$\end{tabular}} &
\colhead{\begin{tabular}{@{}c@{}}Assumed X-ray\\[-0.4ex]
origin\end{tabular}} &
\colhead{\begin{tabular}{@{}c@{}}Inferred accretion rate\\[-0.4ex]
\scriptsize $(M_\odot\,{\rm yr^{-1}})$\end{tabular}} &
\colhead{Physical implication}
}
\startdata
\parbox[t]{0.13\textwidth}{LLAGNs} & \parbox[t]{0.18\textwidth}{$ 10^{37}$--$10^{43}$ \citep{2002ApJS..139....1T,2025MNRAS.540.3827A}} & ADAF & $  10^{-3}$--$1$ & \parbox[t]{0.22\textwidth}{$\dot M$ is consistent with a radiatively inefficient flow.} \\
 &  & disk/corona & $ 10^{-5.5}$--$10^{-2.5}$ & \parbox[t]{0.22\textwidth}{$\dot M$ is too low to support a radiatively efficient disk.} \\
\hline
\parbox[t]{0.13\textwidth}{Seyfert galaxies and radio-quiet quasars} & \parbox[t]{0.18\textwidth}{$ 10^{42}$--$10^{45}$ \citep{1978MNRAS.183..129E,2011AA...533A.128S}} & ADAF & $ 10^{-0.5}$--$10$ & \parbox[t]{0.22\textwidth}{$\dot M$ is too high to be sustained in the ADAF regime.} \\
 &  & disk/corona & $10^{-3}$--$10^{-1.5}$ & \parbox[t]{0.22\textwidth}{This $\dot M$ is more naturally matched to a radiatively efficient flow.} \\
\hline
\parbox[t]{0.15\textwidth}{Jetted AGNs} & \parbox[t]{0.18\textwidth}{$ 10^{40}$--$10^{45}$ \citep{2022PASJ...74..791K}} & disk/corona & $ 10^{-4}$--$10^{1.5}$ & \parbox[t]{0.22\textwidth}{$D$ cannot be disentangled from $\dot M$.} \\
 &  & EC-BLR jet & $ 10^{-2}$--$10^{0.5}$ &  \\
 &  & SSC jet & $ 10^{-1.5}$--$10^{1}$ &  \\
\hline
\parbox[t]{0.11\textwidth}{Some extreme blazars} & \parbox[t]{0.18\textwidth}{$\gtrsim 10^{47}$ \citep{2015JHEAp...7..163G}} & EC-BLR jet & $\gtrsim 10^{1.5}$ & \parbox[t]{0.22\textwidth}{The brightest sources are likely strongly Doppler affected.} \\
 &  & SSC jet & $ \gtrsim 10^{2}$ &  \\
\enddata
\end{deluxetable*}

The X-ray emission is usually dominated by a single component. For non-jetted AGNs, low-luminosity AGNs (LLAGNs) and Seyfert galaxies/radio-quiet quasars exhibit systematically different X-ray luminosities consistent with different accretion states. Their observed X-ray luminosities and the corresponding inferred ranges of the accretion rate are therefore listed separately in Table~\ref{tab:x_mdot}. A comparison of these inferred accretion rates with the physically expected ranges for different accretion modes shows that $\dot M \lesssim 10^{-2.5}\ M_\odot \,\mathrm{yr^{-1}}$ is not physically plausible for a radiatively efficient flow. Accretion rates as high as $\dot M \gtrsim 10^{-0.5}\ M_\odot \,\mathrm{yr^{-1}}$ are also difficult to sustain in an ADAF. This suggests that the X-ray emission of LLAGNs may originate from an ADAF, while those of Seyfert galaxies and radio-quiet quasars are more likely to originate from a radiatively efficient disk or corona.

\section{A measure of radio loudness: Radio-to-X-ray luminosity ratio\label{sec5}}

\begin{deluxetable*}{ll}
\tabletypesize{\normalsize}
\tablewidth{\textwidth}
\tablecaption{Expressions of radio loudness for emission with different origins. Column 1 lists the assumed origins of radio and X-ray emission, and Column 2 gives the corresponding scalings of radio loudness.\label{tab:rxcases}}
\tablehead{\colhead{
Origins of X-ray + Radio }& \colhead{Resulting Radio Loudness }  }
\startdata
\\[-11pt]ADAF + accretion flow &  $4.70 \times 10^{-3} \left(\frac{1+z}{D}\right)^{-\frac{1}{2}} \left(\frac{B}{0.1 \ \mathrm{G}}\right)^{\frac{1}{2}} \left(\frac{\eta}{0.05}\right)^{-1}  \left(\frac{R}{10^2 \  R_{\rm g}}\right)^{\frac{1}{2}}     \left(\frac{\dot{M}}{M_\odot \rm yr^{-1}}\right)^{-1}  \left(\frac{M_{\rm BH}}{10^8 M_\odot }\right) \left(\frac{\nu_{\rm c,syn,obs}}{10^{14} \ \mathrm{Hz}}\right)^{-\frac{1}{2}}$  \\[8pt]
disk + accretion flow &  $2.29 \times 10^{-7} \left(\frac{B}{0.1 \ \mathrm{G}}\right)^{2} \left( \frac{\zeta}{0.1} \right)^{-1}    \left(\frac{R}{10^2 \  R_{\rm g}}\right)^{2}     \left(\frac{\dot{M}}{M_\odot \rm yr^{-1}}\right)^{-1}  \left(\frac{M_{\rm BH}}{10^8 M_\odot }\right)^2$ \\[8pt]
ADAF + weak outflow & $6.66 \times 10^{-7} \left(\frac{1+z}{D}\right)^{-\frac{1}{2}} \left(
\frac{\Gamma}{1}
\right)^{-1} \left(\frac{B}{0.1 \ \mathrm{G}}\right)^{\frac{1}{2}} \left(\frac{\eta}{0.05}\right)^{-2} \left(
\frac{\mathcal{F}}{10^{-4}}
\right) \left(\frac{\dot{M}}{M_\odot \rm yr^{-1}}\right)^{-1}  \left(\frac{M_{\rm BH}}{10^8 M_\odot }\right) \left(\frac{\nu_{\rm c,syn,obs}}{10^{14} \ \mathrm{Hz}}\right)^{-\frac{1}{2}}$  \\[8pt]
disk + weak outflow & $3.24 \times 10^{-11}  \left(
\frac{\Gamma}{1}
\right)^{-1} \left(\frac{B}{0.1 \ \mathrm{G}}\right)^{2} \left(\frac{\eta}{0.05}\right)^{-1} \left(
\frac{\mathcal{F}}{10^{-4}}
\right)\left( \frac{\zeta}{0.1} \right)^{-1}\left(\frac{R}{10^2 \  R_{\rm g}}\right)^{\frac{3}{2}} \left(\frac{\dot{M}}{M_\odot \rm yr^{-1}}\right)^{-1}  \left(\frac{M_{\rm BH}}{10^8 M_\odot }\right)^2 $  \\[8pt]
ADAF + jet & $3.33 \times 10^{-5}  \left(\frac{1+z}{D}\right)^{\frac{1}{2}} \left(
\frac{\Gamma}{20}
\right)^{-1} \left(\frac{B}{0.1 \ \mathrm{G}}\right)^{\frac{1}{2}} \left(\frac{\eta}{0.05}\right)^{-2} \left(
\frac{\mathcal{F}}{0.1}
\right)
\left(\frac{R'}{10^4 \  R_{\rm g}}\right) \left(\frac{R}{10 \  R_{\rm g}}\right)^{-1}
\left(\frac{\dot{M}}{M_\odot \rm yr^{-1}}\right)^{-1}  \left(\frac{M_{\rm BH}}{10^8 M_\odot }\right) \left(\frac{\nu_{\rm c,syn,obs}}{10^{18} \ \mathrm{Hz}}\right)^{-\frac{1}{2}}$  \\[8pt]
disk + jet & $1.62 \times 10^{-7}  \left(\frac{1+z}{D}\right)^{ \frac{1}{2}}  \left(\frac{\Gamma}{20}
\right)^{-1} \left( \frac{\eta}{0.05} \right)^{-1}   \left(\frac{B}{0.1 \ \mathrm{G}}\right)^{2}  \left(\frac{R'}{10^4 \  R_{\rm g}}\right) \left(\frac{R}{10^2 \  R_{\rm g}}\right)^{-\frac{1}{2}} \left(\frac{\dot{M}}{M_\odot \rm yr^{-1} }\right)^{-1} \left(\frac{M_{\rm BH}}{10^8 M_\odot }\right)^2 $  \\[8pt]
jet + jet (SSC) & $6.67\times 10^{-1} \, \left(\frac{1+z}{D}\right)^{-\frac{1}{2}} \left(
\frac{\Gamma}{20}
\right) \left(\frac{B'}{0.1 \ \mathrm{G}}\right)^{\frac{1}{2}} \left(
\frac{\mathcal{F}}{0.1}
\right)^{-1}  \left(\frac{R'}{10^4 \  R_{\rm g}}\right)  \left(\frac{\dot{M}}{M_\odot \rm yr^{-1}}\right)^{-1} \left(\frac{M_{\rm BH}}{10^8 M_\odot }\right) \left(\frac{\nu_{\rm c,syn,obs}}{10^{18} \ \mathrm{Hz}}\right)^{ -\frac{1}{2}}$   \\[8pt]
jet + jet (EC) &  $2.29 \times 10^{-2} \left( \frac{\xi}{0.1} \right)^{-1} \left( \frac{\zeta}{0.1} \right)^{-1}   \left(\frac{B'}{0.1 \ \mathrm{G}}\right)^{2}  \left(\frac{R'}{10^4 \  R_{\rm g}}\right)^{2} \left(\frac{\dot{M}}{M_\odot \rm yr^{-1} }\right)^{-1} \left(\frac{M_{\rm BH}}{10^8 M_\odot }\right)^2$  \\[6pt]
\enddata
\end{deluxetable*}

Motivated by the radio--X-ray correlation, the radio loudness 
\begin{equation}
R_{\rm X} = \frac{L_{\rm R}}{L_{\rm X}},
\end{equation}
can be used to classify AGNs, and a common boundary is $\log R_{\rm X} = -2.755 \pm 0.015$ \citep{2003ApJ...583..145T,2016A&ARv..24...13P,2007A&A...467..519P,2022MNRAS.515..473P}.

\begin{figure*}[ht!]
\centering
\includegraphics[width=0.98\textwidth]{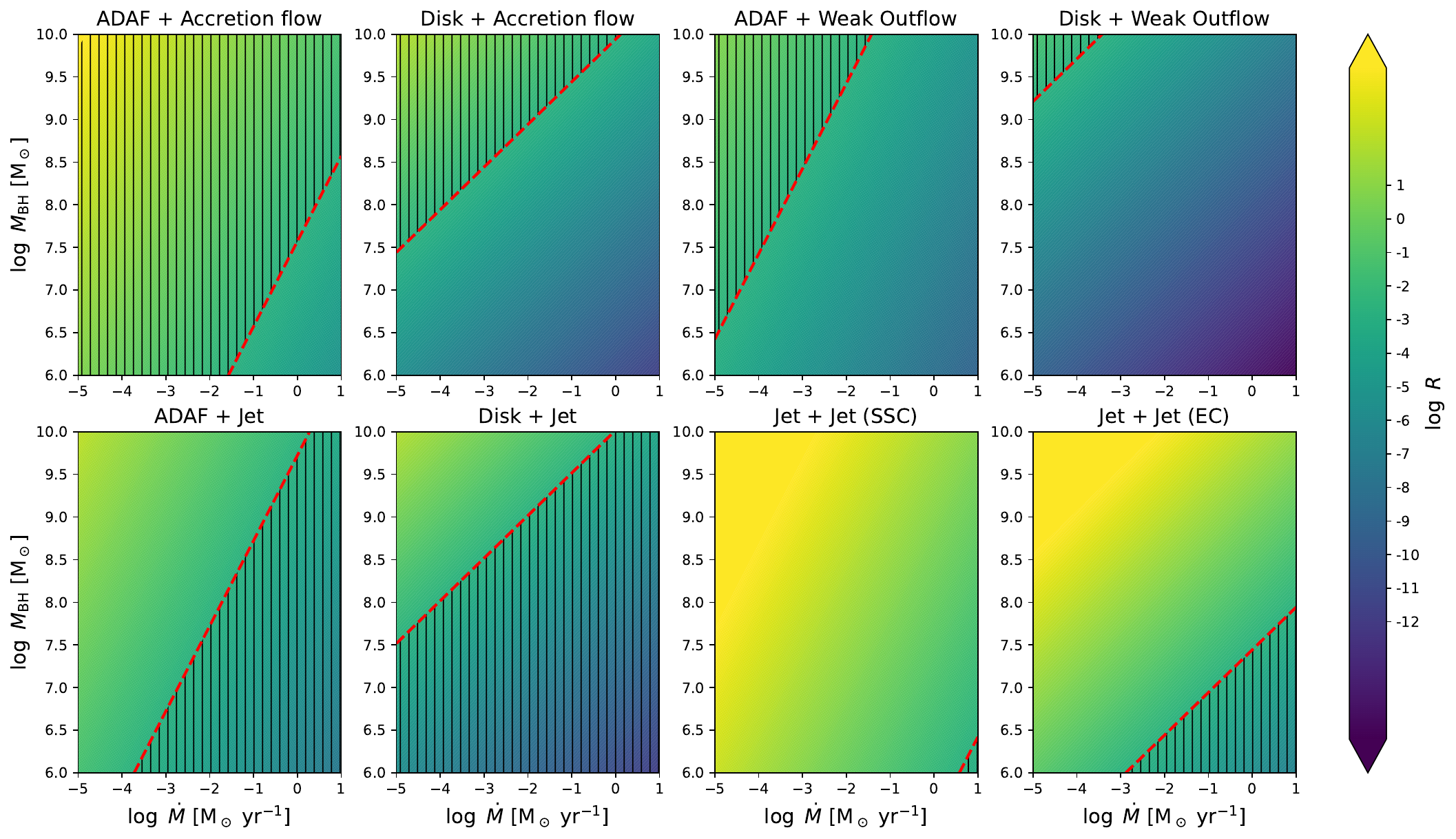}
\caption{Radio loudness over a range of BH masses and mass accretion rates. The upper four panels show the results for non-jetted AGNs, whereas the lower four panels show those for jetted AGNs. Non-jetted AGNs are expected to lie below this threshold, whereas jetted AGNs are expected to lie above it. The shaded regions represent the excluded parameter space. \label{fig33}}
\end{figure*}

We use the radio loudness to further constrain the viable emission configurations. The representative radio/X-ray combinations considered below are listed in Table~\ref{tab:rxcases}. The magnetic field strength, the location and size of the emission region, synchrotron peak frequency, BH mass, and accretion rate all influence radio loudness. The BH mass and the accretion rate span the widest ranges in the parameter space, and therefore largely determine its overall trends. We use $\log R_{\rm X,crit}\simeq -2.755$ as the threshold for radio loudness \citep{2007A&A...467..519P}, and the resulting constraints over typical ranges of $\dot M\simeq 10^{-5}-10^{1}\,M_\odot\,{\rm yr^{-1}}$ \citep{2008MNRAS.390..847L,2018MNRAS.478..399B,2021ApJ...910L..13E} and $M_{\rm BH}\simeq 10^6-10^{10}\,M_\odot$ are shown in Figure~\ref{fig33}.

As shown in Figure~\ref{fig33}, we regard configurations that satisfy the requirements for radio loudness only under extreme parameter choices as unfavorable within the parameter space considered here. For non-jetted AGNs, we interpret the X-rays in LLAGNs as arising from an ADAF, while the radio emission is attributed to a weak outflow. For Seyfert galaxies, the radio emission may arise from either the accretion flow or an outflow because the X-rays are more likely to originate in the disk. For jetted AGNs, radio and X-ray emission are more naturally explained by a jet-dominated scenario, whereas scenarios in which X-rays are produced by non-thermal electrons in the disk or corona are less favorable.

As shown in Table~\ref{tab:rxcases}, we have $R_{\rm X}\propto \dot{M}^{-1} M_{\rm BH}$ or $R_{\rm X}\propto \dot{M}^{-1} M_{\rm BH}^{2}$. These relations imply that the radio loudness increases with the BH mass, which is consistent with the observations reported by \citet{2018ApJ...860..134Q}. By using the Eddington ratio $\lambda_{\rm Edd}\propto \dot{M} \, M_{\rm BH}^{-1}$, we can rewrite $R_{\rm X}\propto \lambda_{\rm Edd}^{-1}$ or $R_{\rm X}\propto M_{\rm BH}\,\lambda_{\rm Edd}^{-1}$, which recovers the anti-correlation between the radio loudness and the Eddington ratio \citep{2002ApJ...564..120H,2007A&A...467..519P,2007ApJ...658..815S,2024A&A...689A.327W}. That is, a lower $\dot M$ (or lower $\lambda_{\rm Edd}$) corresponds to a higher radio loudness.

A widely accepted interpretation is that low accretion rates tend to produce hot flows, such as ADAFs, where radiative cooling is inefficient \citep{2014ARA&A..52..529Y}. The magnetic field can be amplified through dynamo processes, inward advection, or the accumulation of a turbulent magnetic flux. This leads to an enhanced magnetic pressure and the formation of magnetically dominated regions. Such amplification is widely regarded as a key ingredient for launching jets, as it supports the development of large-scale poloidal fields and vertical outflows \citep{2020MNRAS.494.3656L}. This is consistent with the magnetically arrested disk (MAD) state, where the accumulation of magnetic flux in low-$\dot{M}$ environments supports the production of powerful relativistic jets \citep{2011MNRAS.418L..79T,2022MNRAS.511.3795N}. Consequently, systems with low $\dot{M}$ values may naturally evolve toward conditions that are favorable for jet production and enhanced radio emission, which is consistent with our model.

The proposed model has certain limitations. In particular, the actual scale and geometry of the emission region may vary significantly and are difficult to observationally constrain. Moreover, our model does not account for temporal variability or changes in the geometry of the external radiation field, nor does it consider the radial evolution of the magnetic field with distance from the BH. It also uses simplified choices for the peak and cutoff frequencies. Nevertheless, the proposed model suggests that a larger $M_{\rm BH}$ and a lower $\dot{M}$ favor the generation of jets or outflows, which is consistent with observational statistics. The same trend appears in all cases considered here, which suggests that the trends of increasing radio loudness with BH mass and decreasing radio loudness with the Eddington ratio are robust.

\section{Fundamental plane of BH activity \label{sec6}}

The central supermassive BH also plays a crucial role in generating multiwavelength emission. The well-known FP, which relates the radio luminosity ($L_{\rm R}$), X-ray luminosity ($L_{\rm X}$), and BH mass ($M_{\rm BH}$), has been observationally established in several studies \citep{2003MNRAS.345.1057M,2007A&A...467..519P,2022MNRAS.515..473P,2025ApJ...980..187L,2025arXiv250408067K}. It can be written as:
\begin{equation}
\log L_{\rm R} = \xi_{\rm X}\,\log L_{\rm X} + \xi_{\rm M}\,\log M_{\rm BH} + C,
\label{eq:fp}
\end{equation}
where $C$ is a constant, and we refer to $\xi_{\rm X}$ as the slope of the FP.

Observational and theoretical studies suggest that non-jetted and jetted AGNs follow different FP slopes \citep{2006ApJ...645..890W,2003ApJ...583..145T,2008ApJ...688..826L,2022MNRAS.513.4673B,2024A&A...689A.327W}. Although Doppler boosting affects the radio emission of jetted AGNs, this is still not sufficient to explain the observed radio luminosity of these sources \citep{2008ApJ...688..826L}. This suggests that the FP depends on the underlying accretion state, and can be discussed in three broad classes:
(1) radiatively efficient accretion flows (e.g., standard thin disks) \citep{2003MNRAS.345.1057M,2014ApJ...787L..20D,2015MNRAS.447.1289P,2024Univ...10..335Y}, (2) radiatively inefficient accretion flows (e.g., ADAFs) \citep{2003MNRAS.343L..59H, 2003MNRAS.345.1057M,2004MNRAS.355..835H}, and (3) jet-dominated flows \citep{2004MNRAS.355..835H, 2005ApJ...629..408Y, 2009ApJ...703.1034Y,2020MNRAS.497..482L,2025ApJ...980..187L}.

The FP slope $\xi_{\rm X}$ is steeper in radiatively efficient flows than in radiatively inefficient flows, and in jetted AGNs than in non-jetted AGNs \citep{2006ApJ...645..890W,2008ApJ...688..826L,2009ApJ...703.1034Y,2014ApJ...787L..20D,2022MNRAS.513.4673B,2024A&A...689A.327W}. The prevailing view is that changes in the Eddington ratio alter the accretion structure, and thereby drive the observed differences among AGN classes \citep{2002ApJ...564..120H,2007A&A...467..519P,2007ApJ...658..815S,2024A&A...689A.327W}. However, this view relies mainly on statistical trends. We therefore use this framework to probe the underlying physical driver of $\xi_{\rm X}$ and explore one physically plausible interpretation.

Using Equation~\ref{eq:fp}, we express the FP slopes at fixed $M_{\rm BH}$ and fixed $L_{\rm X}$ as logarithmic derivatives,
\begin{widetext}
\begin{equation}
\begin{aligned}
\left.\xi_{\rm X}\right|_{M_{\rm BH}}
&=
\left.
\frac{\partial \log L_{\rm R}}
{\partial \log L_{\rm X}}
\right|_{M_{\rm BH}}=
\frac{
\left.
\dfrac{\partial \log L_{\rm R}}
{\partial \log \dot M}
\right|_{M_{\rm BH}}
}{
\left.
\dfrac{\partial \log L_{\rm X}}
{\partial \log \dot M}
\right|_{M_{\rm BH}}
}, \ \ \ \ \ \  \ \ \ \ 
\left.\xi_{\rm M}\right|_{L_{\rm X}}
&=
\left.
\frac{\partial \log L_{\rm R}}
{\partial \log M_{\rm BH}}
\right|_{L_{\rm X}}
=
\left.
\frac{\partial \log L_{\rm R}}
{\partial \log M_{\rm BH}}
\right|_{\dot M}
-
\left.\xi_{\rm X}\right|_{M_{\rm BH}}
\left.
\frac{\partial \log L_{\rm X}}
{\partial \log M_{\rm BH}}
\right|_{\dot M}.
\end{aligned}
\label{eq:xiM_chain}
\end{equation}
\end{widetext}

The slopes can be calculated directly from the luminosity in our model. For the ADAF case, we use the accretion flow radio luminosity (Equation~\ref{eq:1}) and the ADAF X-ray luminosity (Equation~\ref{eq:3}). For the disk case, we use the accretion flow radio luminosity (Equation~\ref{eq:1}) and the EC-disk X-ray luminosity (Equation~\ref{eqx}). We allow both $B$ and $R$ to depend on $\dot M$ and $M_{\rm BH}$, which gives

{\setlength{\abovedisplayskip}{7pt}
\begin{widetext}
\begin{equation}
\begin{aligned}
\text{ADAF case\label{EQ1}}\quad
\left\{
\begin{aligned}
\left.\xi_{\rm X}\right|_{M_{\rm BH}}
&
=
\frac{\,1+\tfrac{3}{2}\left.\dfrac{\partial \log B}{\partial \log \dot M}\right|_{M_{\rm BH}}
        +\tfrac{3}{2}\left.\dfrac{\partial \log R}{\partial \log \dot M}\right|_{M_{\rm BH}}
        \,}
     {\,2+\left.\dfrac{\partial \log B}{\partial \log \dot M}\right|_{M_{\rm BH}}
        +\left.\dfrac{\partial \log R}{\partial \log \dot M}\right|_{M_{\rm BH}}
\,} \\[8pt]
\left.\xi_{\rm M}\right|_{L_{\rm X}}
&=
\Big[1+\tfrac{3}{2}\left.\dfrac{\partial \log B}{\partial \log M_{\rm BH}}\right|_{\dot M}
      +\tfrac{3}{2}\left.\dfrac{\partial \log R}{\partial \log M_{\rm BH}}\right|_{\dot M}
      \Big]
-\left.\xi_{\rm X}\right|_{M_{\rm BH}}
 \Big[\left.\dfrac{\partial \log B}{\partial \log M_{\rm BH}}\right|_{\dot M}
     +\left.\dfrac{\partial \log R}{\partial \log M_{\rm BH}}\right|_{\dot M}
\Big],
\end{aligned}
\right.
\end{aligned}
\end{equation}

\begin{equation}
\begin{aligned}
\text{Disk case\label{EQ2}}\quad
\left\{
\begin{aligned}
\left.\xi_{\rm X}\right|_{M_{\rm BH}}
&=
\frac{\,1+\tfrac{3}{2}\left.\dfrac{\partial \log B}{\partial \log \dot M}\right|_{M_{\rm BH}}
        +\tfrac{3}{2}\left.\dfrac{\partial \log R}{\partial \log \dot M}\right|_{M_{\rm BH}}
        \,}
     {\,2-\tfrac{1}{2}\left.\dfrac{\partial \log B}{\partial \log \dot M}\right|_{M_{\rm BH}}
        -\tfrac{1}{2}\left.\dfrac{\partial \log R}{\partial \log \dot M}\right|_{M_{\rm BH}}
\,} \\[8pt]
\left.\xi_{\rm M}\right|_{L_{\rm X}}
&=
\Big[1+\tfrac{3}{2}\left.\dfrac{\partial \log B}{\partial \log M_{\rm BH}}\right|_{\dot M}
      +\tfrac{3}{2}\left.\dfrac{\partial \log R}{\partial \log M_{\rm BH}}\right|_{\dot M}
\Big]
-\left.\xi_{\rm X}\right|_{M_{\rm BH}}
 \Big[-1-\tfrac{1}{2}\left.\dfrac{\partial \log B}{\partial \log M_{\rm BH}}\right|_{\dot M}
        -\tfrac{1}{2}\left.\dfrac{\partial \log R}{\partial \log M_{\rm BH}}\right|_{\dot M}
\Big].
\end{aligned}
\right.
\end{aligned}
\end{equation}
\end{widetext}}

\begin{figure*}[ht!]
\centering
\includegraphics[width= 0.9\textwidth]{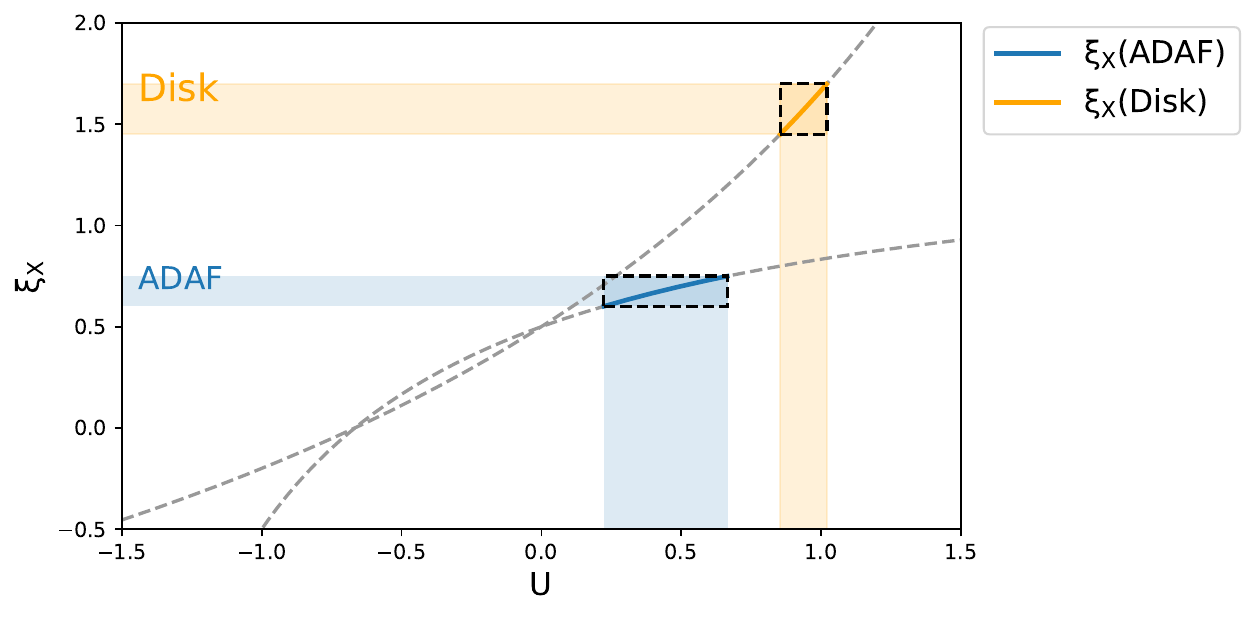}
\caption{Relation between the FP slope $\xi_{\rm X}$ and $U$ given by Equations~\ref{EQ1} and \ref{EQ2}. The range of the FP slope inferred from the sample analysis constrains the parameter space. Only values of $U$ that reproduce the observed range of slopes are regarded as physically plausible. The dashed lines represent the model relations outside the observed range of the slope, and are therefore excluded. The allowed ranges for the ADAF and disk cases are shown by the blue and orange lines, respectively.
\label{fig3}}
\end{figure*}

We make three assumptions for analytical tractability: (1) The magnetic field strength and emission-region radius are most sensitive to changes in $\dot M$, and can be treated as the dominant drivers of $\xi_{\rm X}$. (2) The BH mass has little effect on how the magnetic field strength and the emission-region radius vary with the mass accretion rate. (3) Doppler effects on the FP can be neglected. Under these assumptions, the FP slope $\xi_{\rm X}$ is determined primarily by $U$, where
\begin{equation}U=\dfrac{\partial \log B}{\partial \log \dot M}+ \dfrac{\partial \log R}{\partial \log \dot M}.
\end{equation}

We compare our slopes with predictions reported by \citet{2003MNRAS.343L..59H,2003MNRAS.345.1057M}. They parameterized the slope in terms of the electron spectral index $p$ and the radio spectral index $\alpha$, both of which are reasonably well constrained observationally. We therefore use $p=2$ and $\alpha= 0-0.5$ to estimate the expected range of the slope \citep{2022MNRAS.513.4673B,2024A&A...689A.327W,2025ApJ...980..187L}, which yields $\xi_{\rm X}=0.6-0.75$ for the ADAF and $\xi_{\rm X}=1.45-1.7$ for the disk. Figure~\ref{fig3} shows $\xi_{\rm X}$ as a function of $U$. A smaller $U$ ($0.22-0.67$) implies lower accretion and an ADAF regime, while a larger $U$ ($0.85-1.02$) implies higher accretion and a disk regime.

To interpret the allowed range of $U$, we follow the framework of \citet{2003MNRAS.345.1057M}, in which a fraction of the gravitational power is dissipated in the corona or the ADAF.

In the disk/corona case, the magnetic energy density follows $B^2 \propto  \dot M $ as long as gas pressure dominates, and this leads to $\frac{\partial \log B}{\partial \log \dot M} \simeq 0.5$. Given that radiation pressure-dominated regions of the disk further reduce this derivative, we expect $\frac{\partial \log R}{\partial \log \dot M} > 0$. A plausible interpretation is that the characteristic size of the emission region increases with $\dot M$.

In the ADAF case, the magnetic energy density follows $B^2 \propto \dot M^{3.4}$. This gives $\frac{\partial \log B}{\partial \log \dot M} \simeq 1.7$. We therefore expect $\frac{\partial \log R}{\partial \log \dot M} < 0$. As $\dot M$ increases, the hot inner flow may contract, which in turn causes the emission region to shrink.

We infer that as $\dot M$ increases, the characteristic scale of the emission region in the ADAF decreases, while that in the disk increases. This suggests a transition toward a more radiatively efficient flow. \citet{2000A&A...354L..67M,2017xru..conf..184Q, 2014ARA&A..52..529Y,2022iSci...25j3544L} discussed how the characteristic scale changes with $\dot M$, which is consistent with this inference. We further argue that the differences in slope reflect differences in how the key physical parameters scale with $\dot M$ in different accretion structures. The Eddington ratio affects the accretion structure, which in turn shapes the AGN class. We emphasize that this is only one plausible explanation, and testing it in detail requires MHD simulations. This is beyond the scope of this work.

\section{Conclusion and Discussion \label{sec7}}

We have explored a shock-based framework in which non-thermal electrons accelerated in or near the emission region produce radio and X-ray emission in AGNs. In this scenario, the radio emission is attributed to synchrotron radiation from non-thermal electrons in the accretion flow, jet, or weak outflow, while the X-ray emission is attributed to IC scattering by the same electron population. The intrinsic radio luminosity scales as $L_{\rm R} \propto \dot M M_{\rm BH}$, which is consistent with the empirical trend reported by \citet{2001ApJ...551L..17L}. The X-ray scaling depends on the dominant seed photon field: For the ADAF and SSC-jet cases, we obtain $L_{\rm X} \propto \dot{M}^2$, whereas for the disk/corona and EC-jet cases, we obtain $L_{\rm X} \propto \dot{M}^2 M_{\rm BH}^{-1}$. The additional mass dependence in the disk/corona and EC-jet case arises because the energy density of the external seed photon field scales as $M_{\rm BH}^{-1}$. By applying this framework to non-jetted AGNs, we find that the inferred ranges in the accretion rate favor an ADAF origin for X-ray emission in LLAGNs, whereas Seyfert galaxies and radio-quiet quasars are more naturally associated with radiatively efficient flows. Overall, these results suggest that the main radio/X-ray behavior of typical non-jetted AGNs can be understood within a single shock-based picture, although the preferred emission region still depends on the accretion state.

The observed radio loudness further constrains the viable combinations of radio and X-ray emission sites. In LLAGNs, it favors an ADAF origin for the X-rays together with radio emission from a weak outflow, whereas the X-rays are more naturally associated with the disk in Seyfert galaxies, while radio emission remains compatible with either accretion flow or an outflow. In jetted AGNs, the radio loudness is more naturally matched by jet-dominated scenarios, while disk/corona-based non-thermal X-ray scenarios are less favored. Within this framework, the trend that radio loudness increases with the BH mass and decreases with the Eddington ratio is also consistent with observations reported by \citet{2018ApJ...860..134Q,2002ApJ...564..120H,2007ApJ...658..815S}. Thus, the radio loudness is useful not only as a classification metric, but also as a constraint on where the dominant non-thermal emission is likely to emerge in different AGN classes.

The comparison between the modeled and observed slopes of the FP suggests that the magnetic field strength and the emission-region radius respond differently to changes in the accretion rate in the ADAF and disk-dominated systems. In the disk case, the emission-region radius increases with $\dot M$, whereas it decreases in the ADAF case. This implies that as the accretion rate increases, the ADAF region contracts and the disk extends inward, indicating a transition toward a more radiatively efficient flow. The accretion rate or, equivalently, the Eddington ratio, therefore helps determine the accretion mode, and the resulting accretion structure in turn helps shape the observed AGN class. Under this interpretation, the FP slope does not simply represent an empirical correlation, but also contains information about how the characteristic magnetic field strength and emission-region radius change across accretion states. At the same time, this framework does not provide a satisfactory interpretation of the FP in jetted AGNs.

This work has several limitations. The effects of the mass accretion rate and Doppler boosting cannot be cleanly disentangled, especially in jetted systems. The disk-dominated case also cannot be clearly separated from the corona-dominated case when both are described by non-thermal electrons. Our results should therefore be viewed as a physically motivated framework, rather than as a unique source model. Nevertheless, the main trends of radio and X-ray emission in non-jetted AGNs can be interpreted within this framework, where a single population of shock-accelerated non-thermal electrons produces radio synchrotron emission and X-ray IC emission in different emission regions. The same picture is also broadly consistent with jet-dominated interpretations of jetted AGNs, although it does not by itself replace full SED modeling. A fuller treatment of the geometry, relativistic beaming, and broadband radiative coupling will be needed in future work.

\begin{acknowledgments}
We are very grateful to the anonymous reviewer for the valuable suggestions and comments, which significantly improved the quality of the paper. We would like to express our special thanks to an anonymous colleague and friend for extensive and valuable discussions and assistance. This work was partially supported by the National Science Foundation of China (12263007 and 12233006), and by the High-level Talent Support Program of Yunnan Province.
\end{acknowledgments}

\appendix

\section{Synchrotron luminosity\label{app:1}}

Radio emission is generally attributed to synchrotron radiation, for which the power radiated by a single electron is given by \citep{1979rpa..book.....R}
\begin{equation}
P_{\rm syn}(\gamma)=\frac{4}{3}c \sigma_T \gamma^2 U_{\rm B},
\end{equation}
where $U_{\rm B}=\frac{B^2}{8\pi}$ is the magnetic energy density.

In the observer's frame, the characteristic synchrotron frequency of an electron with Lorentz factor $\gamma$ is written as
\begin{equation}
\nu_{\rm syn,obs}(\gamma) \simeq \frac{D}{1+z} \gamma^2 \nu_{\rm g},
\end{equation}
where $\nu_{\rm g} = \frac{eB}{2 \pi m_{\rm e}c} \approx 2.80 \times 10^5 \left(\frac{B}{0.1 \ \rm G}\right)\ \text{Hz}$ is the gyrofrequency.

We define $\nu_{\rm c,syn,obs}\equiv \nu_{\rm syn,obs}(\gamma_{\max})$ as the observed synchrotron cutoff frequency, which depends on the maximum Lorentz factor of the non-thermal electrons $\gamma_{\max}$ and can therefore be used to infer $\gamma_{\max}$ for given source parameters. In practice, the observed cutoff may also be affected by escape, radiative cooling, and multiple emission components. We therefore adopt relatively high fiducial values: $\nu_{\rm c,syn,obs}\simeq 10^{14}\,{\rm Hz}$ for non-jetted AGNs and $\nu_{\rm c,syn,obs}\simeq 10^{18}\,{\rm Hz}$ for jetted AGNs \citep{2000AA...354..453H,2011AA...525A.118I,2019ARAA..57..467B}. This gives
\begin{equation}
\gamma_{\max}\sim 10^4
\left(\frac{1+z}{D}\right)^{\frac{1}{2}}
\left(\frac{B}{0.1\,{\rm G}}\right)^{-\frac{1}{2}}
\left(\frac{\nu_{\rm c,syn,obs}}{10^{14}\,{\rm Hz}}\right)^{\frac{1}{2}},
\label{eq:gamma_max_nonjetted}
\end{equation}
for non-jetted AGNs, and
\begin{equation}
\gamma'_{\max}\sim 10^6
\left(\frac{1+z}{D}\right)^{\frac{1}{2}}
\left(\frac{B'}{0.1\,{\rm G}}\right)^{-\frac{1}{2}}
\left(\frac{\nu_{\rm c,syn,obs}}{10^{18}\,{\rm Hz}}\right)^{\frac{1}{2}},
\label{eq:gamma_max_jetted}
\end{equation}
for jetted AGNs.

Here, we follow the general steps of \citet{2011AA...525A.118I}. Assuming a power-law electron distribution, the total synchrotron power is
\begin{equation}
\begin{aligned}
P_{\rm tot}
&= \int P_{\rm syn}\, N(\gamma)\, d\gamma = \frac{1}{6\pi} \, c\, \sigma_{\rm T} B^2 N_0
\int_{\gamma_{\rm min}}^{\gamma_{\rm max}} \gamma^{2-p} d\gamma \simeq \frac{1}{6\pi} \, c\, \sigma_{\rm T} B^2 N_0 \gamma_{\rm max}.
\label{eq:pp}
\end{aligned}
\end{equation}

Under the assumptions, we use the corresponding $N_0$ and $\gamma_{\max}$ for the accretion flow, jet, and weak outflow cases in Equation~\ref{eq:pp}. This gives

\begin{equation}
\scriptsize
\begin{aligned}
L_{\rm R,flow} &= P_{\rm tot} = \frac{1}{6\pi} \, c\, \sigma_{\rm T} B^2 \left[ 2.62\times10^{54}
\left(\frac{\eta}{0.05}\right)
\left(\frac{\dot M}{M_\odot\,{\rm yr}^{-1}}\right)
\left(\frac{R}{10^2R_{\rm g}}\right)^{3/2}
\left(\frac{M_{\rm BH}}{10^8M_\odot}\right) \right] \times \left[10^4
\left(\frac{1+z}{D}\right)^{\frac{1}{2}}
\left(\frac{B}{0.1\,{\rm G}}\right)^{-\frac{1}{2}}
\left(\frac{\nu_{\rm c,syn,obs}}{10^{14}\,{\rm Hz}}\right)^{\frac{1}{2}}\right] \\
&\simeq
2.77 \times 10^{37}
\left(
\frac{B}{0.1\ \mathrm{G}}
\right)^2
\left(
\frac{\eta}{0.05}
\right)
\left(
\frac{R}{10^2\  R_{\rm g}}
\right)^{\frac{3}{2}}
\left(
\frac{\dot{M}}{M_\odot\ {\rm yr}^{-1}\ }
\right)
\left(
\frac{M_{\rm BH}}{10^8 M_\odot}
\right) \times 
\left[10^4
\left(\frac{1+z}{D}\right)^{\frac{1}{2}}
\left(\frac{B}{0.1\,{\rm G}}\right)^{-\frac{1}{2}}
\left(\frac{\nu_{\rm c,syn,obs}}{10^{14}\,{\rm Hz}}\right)^{\frac{1}{2}}\right]
\\[4pt]
&=
2.77 \times 10^{41}
\left(
\frac{1+z}{D}
\right)^{\frac{1}{2}}
\left(
\frac{B}{0.1\ \mathrm{G}}
\right)^{\frac{3}{2}}
\left(
\frac{\eta}{0.05}
\right)
\left(
\frac{R}{10^2\  R_{\rm g}}
\right)^{\frac{3}{2}}
\left(
\frac{\dot{M}}{M_\odot\ {\rm yr}^{-1}\ }
\right)
\left(
\frac{M_{\rm BH}}{10^8 M_\odot}
\right)
\left(
\frac{\nu_{\rm c,syn,obs}}{10^{14}\ \mathrm{Hz}}
\right)^{\frac{1}{2}} {\rm erg\ s^{-1}},
\end{aligned}
\end{equation}

\begin{equation}
\scriptsize
\begin{aligned}
L'_{\rm R,jet} &= P_{\rm tot} = \frac{1}{6\pi} \, c\, \sigma_{\rm T} B'^2 \left[ 1.86\times10^{54} \left(\frac{\Gamma}{20}\right)^{-1} \left(\frac{\mathcal{F}}{0.1}\right) \left(\frac{\dot M}{M_\odot\,{\rm yr}^{-1}}\right) \left(\frac{R'}{10^4R_{\rm g}}\right) \left(\frac{M_{\rm BH}}{10^8M_\odot}\right) \right] \times \left[10^6
\left(\frac{1+z}{D}\right)^{\frac{1}{2}}
\left(\frac{B'}{0.1\,{\rm G}}\right)^{-\frac{1}{2}}
\left(\frac{\nu_{\rm c,syn,obs}}{10^{18}\,{\rm Hz}}\right)^{\frac{1}{2}}\right] \\
&\simeq
1.96 \times 10^{39}
\left(
\frac{\Gamma}{20}
\right)^{-1}
\left(
\frac{B'}{0.1\ \mathrm{G}}
\right)^2
\left(
\frac{\mathcal{F}}{0.1}
\right)
\left(
\frac{R'}{10^4\  R_{\rm g}}
\right)
\left(
\frac{\dot{M}}{M_\odot\ {\rm yr}^{-1}\ }
\right)
\left(
\frac{M_{\rm BH}}{10^8 M_\odot}
\right)
\times \left[10^6
\left(\frac{1+z}{D}\right)^{\frac{1}{2}}
\left(\frac{B'}{0.1\,{\rm G}}\right)^{-\frac{1}{2}}
\left(\frac{\nu_{\rm c,syn,obs}}{10^{18}\,{\rm Hz}}\right)^{\frac{1}{2}}\right]
\\[4pt]
&=
1.96 \times 10^{43}
\left(
\frac{1+z}{D}
\right)^{\frac{1}{2}}
\left(
\frac{\Gamma}{20}
\right)^{-1}
\left(
\frac{B'}{0.1\ \mathrm{G}}
\right)^{\frac{3}{2}}
\left(
\frac{\mathcal{F}}{0.1}
\right)
\left(
\frac{R'}{10^4\  R_{\rm g}}
\right)
\left(
\frac{\dot{M}}{M_\odot\ {\rm yr}^{-1}\ }
\right)
\left(
\frac{M_{\rm BH}}{10^8 M_\odot}
\right)
\left(
\frac{\nu_{\rm c,syn,obs}}{10^{18}\ \mathrm{Hz}}
\right)^{\frac{1}{2}} {\rm erg\ s^{-1}},
\end{aligned}
\end{equation}

\begin{equation}
\scriptsize
\begin{aligned}
L'_{\rm R,outflow} &= P_{\rm tot} = \frac{1}{6\pi} \, c\, \sigma_{\rm T} B'^2 \left[ 3.71\times10^{50} \left(\frac{\Gamma}{1}\right)^{-1} \left(\frac{\mathcal{F}}{10^{-4}}\right) \left(\frac{\dot M}{M_\odot\,{\rm yr}^{-1}}\right) \left(\frac{R'}{10^2R_{\rm g}}\right) \left(\frac{M_{\rm BH}}{10^8M_\odot}\right) \right] \times \left[10^4
\left(\frac{1+z}{D}\right)^{\frac{1}{2}}
\left(\frac{B'}{0.1\,{\rm G}}\right)^{-\frac{1}{2}}
\left(\frac{\nu_{\rm c,syn,obs}}{10^{14}\,{\rm Hz}}\right)^{\frac{1}{2}}\right] \\
&\simeq
3.92 \times 10^{33}
\left(
\frac{\Gamma}{1}
\right)^{-1}
\left(
\frac{B'}{0.1\ \mathrm{G}}
\right)^2
\left(
\frac{\mathcal{F}}{10^{-4}}
\right)
\left(
\frac{R'}{10^2\  R_{\rm g}}
\right)
\left(
\frac{\dot{M}}{M_\odot\ {\rm yr}^{-1}\ }
\right)
\left(
\frac{M_{\rm BH}}{10^8 M_\odot}
\right)
\times \left[10^4
\left(\frac{1+z}{D}\right)^{\frac{1}{2}}
\left(\frac{B'}{0.1\,{\rm G}}\right)^{-\frac{1}{2}}
\left(\frac{\nu_{\rm c,syn,obs}}{10^{14}\,{\rm Hz}}\right)^{\frac{1}{2}}\right]
\\[4pt]
&=
3.92 \times 10^{37}
\left(
\frac{1+z}{D}
\right)^{\frac{1}{2}}
\left(
\frac{\Gamma}{1}
\right)^{-1}
\left(
\frac{B'}{0.1\ \mathrm{G}}
\right)^{\frac{3}{2}}
\left(
\frac{\mathcal{F}}{10^{-4}}
\right)
\left(
\frac{R'}{10^2 \  R_{\rm g}}
\right)
\left(
\frac{\dot{M}}{M_\odot\ {\rm yr}^{-1}\ }
\right)
\left(
\frac{M_{\rm BH}}{10^8 M_\odot}
\right)
\left(
\frac{\nu_{\rm c,syn,obs}}{10^{14}\ \mathrm{Hz}}
\right)^{\frac{1}{2}} {\rm erg\ s^{-1}}.
\end{aligned}
\end{equation}

The transformation from the comoving frame to the observer's frame is $L_{\rm obs}=\int \frac{D^3 P'}{(1+z)^2} D N'(\gamma) d \gamma =\frac{D^4}{(1+z)^2}L'$. Therefore, the radio luminosities of non-jetted and jetted AGNs in the observer's frame are given by Equations~\ref{eq:1}--\ref{eq:2}.

\section{IC luminosity scalings\label{app:2}}

For an isotropic, low-frequency radiation field, the IC scattering power of a relativistic electron is given by
\begin{equation}
P_{\rm IC}=\frac{4}{3} c\sigma_{\rm T} \gamma^2 U_{\rm ph},
\end{equation}
where $U_{\rm ph}=L_{\rm seed}/(4\pi R^2c)$ is the seed photon energy density, $L_{\rm seed}$ is the seed photon luminosity, and the reference frame in which $L_{\rm seed}$ and $R$ are evaluated depends on the IC process. For the SSC case, the seed photons originate inside the emission region, so $L_{\rm seed}$ and $R$ are evaluated in the comoving frame. For the EC case, the seed photons originate outside the emission region, so $L_{\rm seed}$ and $R$ are evaluated in the AGN rest frame, and $R$ is the distance from the BH to this external region.

We assume that the X-rays are entirely produced by IC scattering. In the Thomson regime,
\begin{equation}
L_{\rm X,obs}  = \frac{D^4}{(1+z)^2} \int P_{\rm IC} N(\gamma) d \gamma  =\frac{D^4}{(1+z)^2}   \frac{1}{3\pi} \sigma_{\rm T} N_0 \gamma'_{\rm max} \frac{L_{\rm  seed}}{R^2}.
\label{eq:IC_general}
\end{equation}

For each case, we use the corresponding $N_0$, $\gamma_{\max}$, $R$, and seed photon field, with all quantities evaluated in the appropriate reference frame. For illustration, the SSC-jet luminosity is

\begin{equation}
\scriptsize
\begin{aligned}
L_{\rm X,SSC,jet,obs}
&=\frac{D^4}{(1+z)^2} \frac{1}{3\pi} \sigma_{\rm T} N_0 \gamma'_{\rm max} \frac{L'_{\rm R,jet}}{R'^2} \\
&=\frac{D^4}{(1+z)^2}  \frac{1}{3\pi} \sigma_{\rm T} \left[ 1.86\times10^{54}
\left(\frac{\Gamma}{20}\right)^{-1}
\left(\frac{\mathcal{F}}{0.1}\right)
\left(\frac{\dot M}{M_\odot\,{\rm yr}^{-1}}\right)
\left(
\frac{R'}{10^4\  R_{\rm g}}
\right)
\left(\frac{M_{\rm BH}}{10^8M_\odot}\right) \right] \times \left[10^6
\left(\frac{1+z}{D}\right)^{\frac{1}{2}}
\left(\frac{B'}{0.1\,{\rm G}}\right)^{-\frac{1}{2}}
\left(\frac{\nu_{\rm c,syn,obs}}{10^{18}\,{\rm Hz}}\right)^{\frac{1}{2}}\right] \\& \times \left[1.96 \times 10^{43}
\left(
\frac{1+z}{D}
\right)^{\frac{1}{2}}
\left(
\frac{\Gamma}{20}
\right)^{-1}
\left(
\frac{B'}{0.1\ \mathrm{G}}
\right)^{\frac{3}{2}}
\left(
\frac{\mathcal{F}}{0.1}
\right)
\left(
\frac{R'}{10^4\  R_{\rm g}}
\right)
\left(
\frac{\dot{M}}{M_\odot\ {\rm yr}^{-1}\ }
\right)
\left(
\frac{M_{\rm BH}}{10^8 M_\odot}
\right)
\left(
\frac{\nu_{\rm c,syn,obs}}{10^{18}\ \mathrm{Hz}}
\right)^{\frac{1}{2}} \right]  \\
&\times \left[ 2.95\times10^{17}\left(
\frac{R'}{10^4\  R_{\rm g}}
\right) \left(
\frac{M_{\rm BH}}{10^8 M_\odot}
\right)\right]^{-2} \\
&= 2.94 \times 10^{43} \left(\frac{D^3}{1+z}\right) \left(
\frac{\Gamma}{20}
\right)^{-2} \left(\frac{B'}{0.1 \ \mathrm{G}}\right) \left(
\frac{\mathcal{F}}{0.1}
\right)^2 \left(\frac{\dot{M}}{M_\odot \rm yr^{-1} }\right)^2
\left(\frac{\nu_{\rm c,syn,obs}}{10^{18} \ \mathrm{Hz}}\right)   \rm erg \ s^{-1}.
\end{aligned}
\end{equation}

The other IC cases are derived analogously, yielding Equations~\ref{eq:3}--\ref{eq:4}.

\bibliography{sample631}{}
\bibliographystyle{aasjournal}
\end{document}